\documentclass[12pt,preprint]{aastex}

\newcommand{\bel}[1]{\begin{equation}\label{#1}}
\newcommand{\be}{\begin{equation}}
\newcommand{\ee}{\end{equation}}
\newcommand{\ba}{\begin{eqnarray}}
\newcommand{\ea}{\end{eqnarray}}
\newcommand{\bal}[1]{\begin{eqnarray}\label{#1}}

\shorttitle{Spins evolution via minor mergers}
\shortauthors{Mandel}

\begin{document}
 
\title{Spin distribution following minor mergers and the effect of spin 
on 
the detection range for low-mass-ratio inspirals}
\author{Ilya Mandel} \affil{Theoretical Astrophysics, California
Institute of Technology, Pasadena, California 
91125}\email{ilya@caltech.edu}


\begin{abstract} 

We compute the probability distribution for the spin of a black hole following
a series of minor mergers with isotropically distributed, non-spinning,
inspiraling compact objects.  By solving the Fokker-Planck equation governing
this stochastic process, we obtain accurate analytical fits for the evolution
of the mean and standard deviation of the spin distribution in several
parameter regimes. We complement these analytical fits with numerical
Monte-Carlo simulations in situations when the Fokker-Planck analysis is not
applicable.  We find that a $\sim 150\ M_\odot$ intermediate-mass black hole
that gained half of its mass through minor mergers with neutron stars will have
dimensionless spin parameter $\chi=a/M \sim 0.2 \pm 0.08$.  We estimate the
effect of the spin of the central black hole on the detection range for 
intermediate-mass-ratio inspiral (IMRI) detections by Advanced LIGO and
extreme-mass-ratio inspiral (EMRI) detections by LISA.  We find that for
realistic black hole spins, the inclination-averaged Advanced-LIGO IMRI
detection range may be increased by up to $10\%$ relative to the range for
IMRIs into non-spinning intermediate-mass black holes.  For LISA, we find that
the detection range for EMRIs into $10^5\ M_\odot$ massive black holes (MBHs)
is not significantly affected by MBH spin, the range for EMRIs into $10^6\
M_\odot$ MBHs is affected at the $\lesssim 10\%$ level, and  EMRIs into
maximally spinning $10^7\ M_\odot$ MBHs are detectable to a distance $\sim 25$
times greater than EMRIs into non-spinning black holes.  The resulting bias in
favor of detecting EMRIs into rapidly spinning MBHs will play a role when
extracting the MBH spin distribution from EMRI statistics.

\end{abstract}


\maketitle

\section{Introduction}

A growing body of evidence from observations, numerical simulations, and
comparisons between the two, suggests the existence of a population of 
intermediate-mass black holes with masses in the $M\sim 10^2-10^4\,M_\odot$
range (e.g., \citep{MC2004} and references therein).  These
intermediate-mass black holes may capture compact objects (stellar-mass black
holes or neutrons stars) and merge with them
\citep{Tan00,MH02a,MH02b,MT02a,MT02b,GMH04,GMH06,OL06, Mandel2007}.  In 
addition
to adding to the black-hole mass, the merging compact objects will also
contribute their orbital angular momentum to the spin angular momentum of the
central black hole, leading to the evolution of the black-hole spin through a
sequence of such minor mergers.  

We might expect the typical spin of a black hole to be low if a significant
fraction of its mass has been added via minor mergers with compact objects
whose angular momentum at plunge is distributed isotropically.  The angular
momentum imparted to the black hole of mass $M$ by a compact object of mass $m$
is $L_{\rm obj} \propto m M$.   (We include only the orbital angular momentum,
not the spin angular momentum of the compact object, since the latter is lower
than the former  by a factor of order $m/M$, which we assume to be small for
minor mergers.) This causes the dimensionless spin parameter of the hole
$\chi\equiv S_1/M^2 = a/M$ to change by $\sim L_{\rm obj}/M^2 \propto m/M$. 
After $N \sim M/m$ such mergers, necessary for the hole to grow to mass $M$,
the typical dimensionless spin parameter of the hole will be $\chi \propto
(m/M)\sqrt{N} \sim \sqrt{m/M}$. 

As discussed by \citet{Miller2002} and \citet{HB}, the
angular momenta of black holes that grow through minor mergers undergo a damped
random walk.  The damping comes about because retrograde orbits, which subtract
angular momentum from a black hole, plunge from a last stable orbit (LSO) at a
higher radius than prograde orbits, so more angular momentum is subtracted
following retrograde inspirals than is added following prograde ones.  

In this paper, we make an analytical approximation to the spin change induced
by a minor merger and solve the Fokker-Planck equation to obtain the evolution
of the spin probability distribution \citep{HB}.  (We use a simpler
one-dimensional version of the Fokker-Planck equation than \citet{HB}, 
since we are interested only in the evolution of the
magnitude of the spin, not its direction.) We find that for black holes with
$\chi \gg \sqrt{m/M}$, the spin $\chi$ evolves proportionally to $M^{-2.63}$ as
the mass grows via minor mergers (rather than $M^{-2}$, which would be the case
without damping).  We determine the asymptotic values of the expected mean  of
the spin distribution and its standard deviation in the limit of infinitely
many minor mergers:  $\bar{\chi}\to \sqrt{1.5 m /M}$ and $\sigma \to \sqrt{0.7
m /M}$.  We also describe the evolution of the spin distribution in other
parameter regimes, e.g., when $\sqrt{m/M} \gg \chi \gg m/M$.

Our Fokker-Planck analysis fails when the mass ratio $m/M$ is not
sufficiently low, so for those cases we resort to Monte-Carlo numerical
simulations.  We find that if the mass of the central black hole grows
from $M=5 m$ to $M=10 m$ by capturing five objects of equal mass $m$, the
mean spin of the resulting black hole is $\bar{\chi} \approx 0.5$, nearly
independent of its initial spin (\citet{Miller2002} obtained similar
results).  However, if the central black hole grows from $M=50 m$ to
$M=100 m$ (e.g., a $M=70\ M_\odot$ black hole growing to $M=140\ M_\odot$
by capturing fifty $m=1.4\ M_\odot$ neutron stars), its resulting spin is
rather low, $\chi \sim 0.2 \pm 0.08$.

The combination of the spin of the central black hole and the inclination of
the inspiraling object's orbit can have a significant effect on the
gravitational-wave signal from a low-mass-ratio inspiral.  We compute the
increase in the Advanced-LIGO detection range for intermediate-mass-ratio
inspirals (IMRIs) due to the spin of the central black hole.  We find that the
detection range, averaged over orbital inclinations, may increase by $\sim
3-10\%$ relative to the range for inspirals into non-spinning black holes for
the expected values of black hole mass and spin.  We provide an approximate
expression for the dependence of the Advanced-LIGO IMRI detection range on spin
[see Eq.~(\ref{AdvLIGORatio})].  We also compute the change in the LISA
extreme-mass-ratio-inspiral (EMRI) detection range due to the spin of the
massive black hole. We find that the range for inspirals into $M=10^5\ M_\odot$
black holes is nearly independent of their spin, because the frequency at the
last stable orbit (LSO) is away from the minimum of the LISA noise curve. On
the other hand, the inclination-averaged detection range for IMRIs into rapidly
spinning $M=10^7\ M_\odot$ black holes is $\sim 25$ times greater than into
non-spinning ones.  The detection volumes are proportional to the cube  of the
range.  This will create a bias in favor of detecting inspirals into
rapidly spinning black holes, which in turn will have consequences for the
extraction of massive-black-hole spin function from LISA EMRI statistics.

The paper is organized as follows.  In Sec.~\ref{SpinEvol}, we provide the
background for our calculation of the spin evolution via minor mergers.  In
Sec.~\ref{FPSpinEvol}, we describe analytical solutions of the Fokker-Planck
equation for spin evolution.  In Sec.~\ref{MCSpinEvol}, we describe
Markov-Chain numerical simulations of spin evolution.  In Sec.~\ref{Boost}, we
evaluate the dependence of the detection ranges for low-mass-ratio inspirals
averaged over orbital inclination angles on the spin of the massive body, in
the context of both Advanced LIGO and LISA.

\section{Spin evolution}\label{SpinEvol}

We assume that the distribution of the orbital inclination angle $\iota$
relative to the central black hole's spin is isotropic at
capture.  Here $\iota$ is defined via
\bel{iota}
\cos{\iota}=\frac{L_z}{\sqrt{L_z^2+Q}},
\ee
$L_z$ is the object's orbital angular momentum in the direction of the black
hole's spin, and $Q$ is the Carter constant.  We further assume that the 
inclination angle $\iota$ remains approximately constant over the
inspiral~\citep{Hughes2000}, so the distribution of inclinations at the 
LSO is also isotropic, ${\rm Pr}(\cos \iota)=1/2$.

In the low-mass-ratio limit, the amount of angular momentum radiated in
gravitational waves during the plunge and ringdown is smaller by a factor of
$\sim m/M$ than the angular momentum at the LSO.  Therefore, we assume that the
merging object contributes its orbital angular momentum at the LSO to the
angular momentum of  the black hole.  The spin of the black hole after a minor
merger, $\chi'$,  is related to the original spin $\chi$ via
\bel{chievol}
\chi' \approx \frac{1}{(M+m)^2} \sqrt{(\chi M^2 + L_z)^2+Q},
\ee
where $m$ is the mass of the small object, $M$ is the mass of the hole, and we
assume $m \ll M$.

The constants of motion $L_z$ and $Q$ at the LSO can be obtained as a function
of $\iota$ by demanding that
the potential $R$ and its first and second derivatives in $r$ are zero at the
LSO (see Chapter 33 of \citep{MTW}):
\bal{R}
\nonumber
R &=& \left[ E (r^2+\chi^2 M^2) - L_z \chi M \right]^2 
	- (r^2-2 M r + \chi^2 M^2)
	\left[m^2 r^2 + (L_z- \chi M E)^2 + Q\right],
	\\
R &=&0, \qquad \frac{dR}{dr} = 0, \qquad \frac{d^2 R}{dr^2}=0 \qquad 
	{\rm at\ LSO}.
\ea

It is possible to make analytic approximations to the values of $L_z$ and $Q$
at the LSO based on appropriately averaging the
analytically known constants of motion at the LSO for prograde and retrograde
equatorial orbits  (cf.~Eq.~(9) of \citep{HB}).  In particular, for $\chi 
\ll
1$, the plunging object's dimensionless ``total angular momentum'' is given by
\bel{Lapprox}
\hat{L}=\frac{\sqrt{L_z^2+Q}}{Mm} \approx M m \sqrt{12} 
	\left[ 1- \frac{1}{2}\left(\frac{2}{3}\right)^{3/2} \chi \cos \iota
	\right],
\ee
where we correct a mistake in Eq.~(4) of \citep{Miller2002}.  Then $L_z$ 
and $Q$ follow from Eq.~(\ref{iota}):
\bel{LzQ}
	L_z = \cos{\iota} \sqrt{L_z^2+Q}; \qquad Q=\sin{\iota} \sqrt{L_z^2+Q} .
\ee

\section{Fokker-Planck equation for spin evolution}\label{FPSpinEvol}

The black-hole spin evolution is a stochastic process.  The probability
distribution function of a stochastic process, however, can be described the
deterministic Fokker-Planck equation:
\bel{FP}
\frac{\partial}{\partial t} f(x,t)=
	-\frac{\partial}{\partial x} \left[\mu(x,t) f(x,t)\right]+ 
	\frac{1}{2}\frac{\partial^2}{\partial x^2} 
		\left[\sigma^2(x,t) f(x,t)\right],
\ee
where $\mu=\langle dx \rangle/dt$ is the mean drift and
$\sigma^2=\langle(dx)^2\rangle/dt$ is the stochastic variance. In this Section, 
we derive approximate analytical solutions to the Fokker-Planck equation in
several interesting parameter regimes.

For simplicity, assume that all merging objects have the same mass $m$.  We
parametrize the mass of the black hole by a dimensionless ``time''
parameter $t=M/m$.  The change in the spin $\chi$ after a merger follows from
Eq.~(\ref{chievol}):
\bel{dchi}
d\chi=\frac{1}{(t+1)^2}\sqrt{\chi^2 t^4 + \hat{L}^2 t^2 + 2\chi \hat{L} t^3
	\cos{\iota}} -\chi.
\ee

We can compute $\hat{L}$ at plunge as a function of $\chi$ and $\cos{\iota}$ by
solving Eqs.~(\ref{R}), then substituting the result into Eq.~(\ref{dchi}) to
obtain $d\chi$ as a function of $t$, $\chi$, and $\cos{\iota}$.  Although this
process is simple in principle, such a numerical computation makes it
impossible to obtain analytic expressions for $\langle d\chi \rangle$ and
$\langle (d\chi)^2 \rangle$, which are necessary if we wish to solve the
Fokker-Planck equation.  (Here, brackets denote averaging over $\cos{\iota}$.)

We could, of course, try to obtain empirical analytic fits to the numerical
solutions for $\langle d\chi \rangle$ and $\langle (d\chi)^2 \rangle$, but it
turns out that there is a simpler approach.  The approximate formula for
$\hat{L}$ given in Eq.~(\ref{Lapprox}) is valid only when $\chi \ll 1$; when
$\chi \sim 1$, Eq.~(\ref{Lapprox}) overestimates $\hat{L}$ by as much as
$40\%$. Remarkably, however, using this incorrect approximation for $\hat{L}$
in Eq.~(\ref{dchi}) generally yields very accurate expressions for $\langle
d\chi \rangle$ for a wide range of $\chi$.  So long as $\chi t \gg 1$ (i.e.,
$\chi \gg m/M$), an expansion of Eq.~(\ref{dchi}) to the first order in
$1/(\chi t)$ yields the following simple analytic expression for the mean drift
in $\chi$:
\bel{mu}
\mu(\chi,t)=\frac{\langle d\chi \rangle}{dt} = 
\frac{\chi}{t} \left(-2-\frac{4\sqrt{2}}{9}\right)+\frac{4}{\chi t^2}. 
\ee 
This expression is accurate to about $1\%$ for all values of $\chi$ so long as
$\chi t \gtrsim 10$.  Similarly, the analytic expression for the stochastic
variance of the spin is 
\bel{sigma} \sigma^2(\chi,t)=\frac{\langle (d\chi)^2 \rangle}{dt}
=\frac{4}{t^2} \left(1+\frac{4 \sqrt{2} \chi^2}{9}-\chi^2\right). 
\ee 
This expression underestimates the variance by $\gtrsim 10\%$ for very  high
spins, but is generally accurate to a few percent for lower spins which are
expected as a consequence of minor mergers in the Advanced LIGO setting.

We can now substitute Eqs.~(\ref{mu}) and (\ref{sigma}) into the Fokker-Planck
equation for the probability evolution (\ref{FP}) to obtain
\bal{FPspin}
\frac{\partial}{\partial t} f(\chi,t)&=&
	-\frac{\partial}{\partial \chi} \left[
	\frac{\chi}{t} \left(-2-\frac{4\sqrt{2}}{9}
		+\frac{4}{\chi^2 t}\right) f(\chi,t)\right]\\
	\nonumber 
	&+& 
	\frac{1}{2}\frac{\partial^2}{\partial \chi^2} 
		\left[\frac{4}{t^2}
	\left(1+\frac{4 \sqrt{2} \chi^2}{9}-\chi^2\right) f(\chi,t)\right].
\ea
This is a one-dimensional equation unlike the three-dimensional equation 
derived in \citep{HB}, since we choose to focus on the evolution of the 
magnitude of the spin, not its direction.  Still, this is a rather complicated 
equation that does not easily separate.  Fortunately, for many applications 
it is not necessary to solve the complete equation.  

Equation (\ref{FPspin}) was derived under the assumption $\chi t \gg 1$.  If we
further assume that $\chi^2 t \gg 1$ (i.e., $\chi \gg \sqrt{m/M}$, 
then the mean spin evolution is dominated by
\bel{DriftSpinApprox}
\frac{d \bar{\chi}}{dt} \approx  a \frac{\bar{\chi}}{t},
\ee
where $a \equiv -2-4\sqrt{2}/9\approx-2.63$.  (This result can also be obtained
directly from Eq.~(\ref{mu}).)  Thus, the mean spin evolves according to
\bel{SpinApprox}
\bar{\chi} \approx \bar{\chi}_0 \left(\frac{t}{t_0}\right)^a
	\approx \bar{\chi}_0 \left(\frac{M_0}{M}\right)^{2.63}
\ee
(compare with Eq.~(26) of \citep{HB}, where the exponent is approximated 
by
$2.4$).

If the assumption $\chi^2 t \gg 1$ is not satisfied, and instead $\chi^2 t \ll
1$, but $\chi t \gg 1$ so that Eq.~(\ref{FPspin}) still holds,  the evolution
of the probability function may be approximated as
\be
\frac{\partial f(t,\chi)}{\partial t}= -\frac{\partial}{\partial \chi}
	\left(\frac{4 f(t,\chi)}{\chi t^2}\right)
	+\frac{1}{2}\frac{\partial^2}{\partial \chi^2}
	\left(\frac{4 f(t,\chi)}{t^2}\right).
\ee
This equation can be solved by separation of variables: $f(t, \chi)=T(t) X(\chi)$,
where the solution for $T$ is $T(t)=\exp(-k/t)$, $X$ is the solution to
\be
2\chi^2 X'' - 4 \chi X + 4 X - k \chi^2 X = 0,
\ee
and $k$ is a constant.  The mean spin grows roughly as
\bel{SpinApprox2}
\bar{\chi} \sim \sqrt{\frac{2}{t_0}-\frac{2}{t}},
\ee
so after $t \gtrsim 2 t_0$ (i.e., after the black hole captures half its mass 
via minor mergers), $\chi^2 t \gtrsim 1$.

The spin growth and spin decay terms in 
Eq.~(\ref{FPspin}) cancel when the spin is approximately equal to 
\bel{SpinAsympt}
\bar{\chi} \to \sqrt{\frac{4}{-a t}} \approx \sqrt{\frac{1.5}{t}}.
\ee
(Compare with \citet{Miller2002}, who estimated the mean spin to be
$\sqrt{2} \sqrt{(m/M)} = \sqrt{2/t}$ based on numerical simulations.)

We can estimate the second moment of the probability distribution by
approximating the solution to Eq.~(\ref{FPspin}) by a Gaussian 
(as suggested by \citet{Miller2002}):
\bel{GaussianGuess}
f(\chi, t)=\frac{1}{\sqrt{2\pi}\sigma} \exp\left[
	-\frac{\left(\chi-\bar{\chi} (t)\right)^2}{2\sigma^2(t)}
		\right].
\ee
(A Gaussian turns out to be a good approximation except at small  
$\bar{\chi}$, when the tails at $\chi>\bar{\chi}$ are larger than those
at $\chi<\bar{\chi}$.)  Substituting this Gaussian into Eq.~(\ref{FPspin}),
keeping only the lowest-order terms in $t\chi$, and setting $\chi =\bar{\chi}$, 
we obtain
\bel{DriftSigmaApprox}
-\frac{1}{\sigma}\frac{d\sigma}{dt} =
	-\frac{a}{t}-\frac{2}{t^2\sigma^2}(1+b\bar{\chi}^2),
\ee
where $b \equiv 4\sqrt{2}/9-1$.
If $\sigma^2 t \gg 1$, then $\sigma$ evolves in the same way as 
$\bar{\chi}$ when $\chi^2 t \gg 1$:
\bel{SigmaApprox}
\sigma \approx \sigma_0 \left(\frac{t}{t_0}\right)^a
	\approx \sigma_0 \left(\frac{M_0}{M}\right)^{2.63}.
\ee 

What if $\sigma^2 t \ll 1$?  This might be the case of interest if, say, the
initial spin of a black hole created during some process is known precisely,
and we wish to estimate future spin evolution through minor mergers.  In this
case, the second term on the right-hand side of Eq.~(\ref{DriftSigmaApprox})
dominates, and if $\bar{\chi}$ is small or does not change
significantly, $\sigma$ grows according to
\bel{SigmaApprox2}
\sigma \approx \sqrt{4 \left(1+b \bar{\chi}^2\right)
	\left(\frac{1}{t_0}-\frac{1}{t}\right)+\sigma_0^2}. 
\ee

In either case, $\sigma$ asymptotes to the solution
\bel{SigmaAsympt}
\sigma \to \sqrt{\frac{2(1+b \bar{\chi}^2)}{-at}}.
\ee
For large $t$, $\sigma \sim \sqrt{2/(-at)} \approx \sqrt{0.7/t}$; 
\citet{Miller2002} estimated $\sigma$ to be
$\sqrt{(m/M)}/\sqrt{2} = \sqrt{1/(2t)}$ based on numerical simulations.

Lastly, consider the case when $\chi t \lesssim 1$.  In this case the orbital
angular momentum of the plunging object is comparable to the spin angular
momentum of the black hole, and Eq.~(\ref{FPspin}) is incorrect, since it was
derived under the assumption $\chi t \gg 1$.  If the black hole is initially
non-spinning or has spin $\chi \lesssim 1/t$, however, a single minor merger
will bring its spin to $\chi \sim \sqrt{12}/t$ according to Eq.~(\ref{dchi}). 
This case can be treated with a Monte-Carlo numerical simulation as described
in the next section.

\section{Spin evolution via Monte Carlo simulations}\label{MCSpinEvol}

We have carried out Monte Carlo simulations of spin evolution through minor
mergers in order to confirm the analytical estimates presented above, based on
the approximate Fokker-Planck equation.  Our simulations also allow us to
access the small-$t$ regime where the Fokker-Planck approach is not valid, but
where our physical approximations for low-mass-ratio inspirals still
hold.  Since these simulations were performed numerically, there was no need
to make analytical approximations to $d\chi$ following a merger; instead, we
solved Eqs.~(\ref{R}) directly and obtained $d\chi$ via Eq.~(\ref{dchi}).

\begin{figure}
\includegraphics[keepaspectratio=true,width=6in]{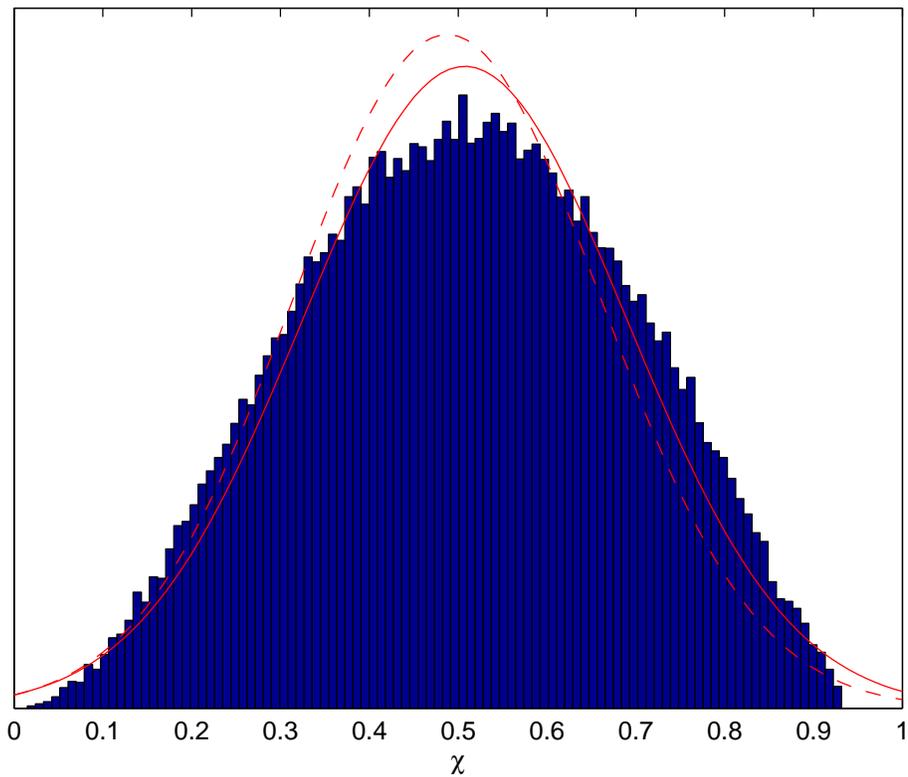}
\caption{Monte-Carlo predictions for the black-hole spin distribution
following black-hole growth via minor mergers from $t=M/m=5$ to $t=M/m=10$.  
The histogram shows the spin distribution at $t=10$ for a black hole with 
initial spin $\chi=0.9$, and the solid curve is a Gaussian fit to that 
distribution.  The dashed curve is a Gaussian fit to the spin distribution at 
$t=10$ for a black hole that has initial spin $\chi=0.1$ at $t=5$.
\label{Hist510}}
\end{figure}

In Figure \ref{Hist510} we plot the spin distribution of a black hole of mass
$t=M/m=10$ that started out with either spin $\chi=0.1$ or $\chi=0.9$ at
$t=M/m=5$ before growing via minor mergers.  This corresponds, for example, to
an intermediate-mass black hole that grows from $M=50\ M_\odot$ to $M=100\
M_\odot$ by capturing $m=10\ M_\odot$ black holes.  The distributions for both
values of initial spin are roughly Gaussian, although with
shorter-than-Gaussian tails (we plot the actual Monte-Carlo histogram for the
$\chi=0.9$ case for comparison with a fitted Gaussian). We see that for these
small values of $t$, the initial value of the spin is largely forgotten after
the black hole captures half of its mass through minor mergers.  The means of
the spin at $t=10$ are $\bar{\chi}=0.49$ for the initially
slowly-spinning hole and $\bar{\chi}=0.51$ for the initially
rapidly-spinning hole.  The standard deviations at $t=10$ are $\sigma=0.17$
for initial spin $\chi=0.1$ and $\sigma=0.18$ for initial spin $\chi=0.9$ (the
initial standard deviations are zero in both cases, i.e., the initial spins are
presumed to be precisely determined).  These results agree with Fig.~1 of
\citep{Miller2002}.  Because the values of $t$ involved are so small, the
Fokker-Planck equation (\ref{FPspin}) does not apply: at $t=5$, the angular
momentum of the inspiraling object at the LSO is comparable to or larger than
the spin angular momentum of the black hole even for large initial black hole
spins.  

\begin{figure}
\includegraphics[keepaspectratio=true,width=6in]{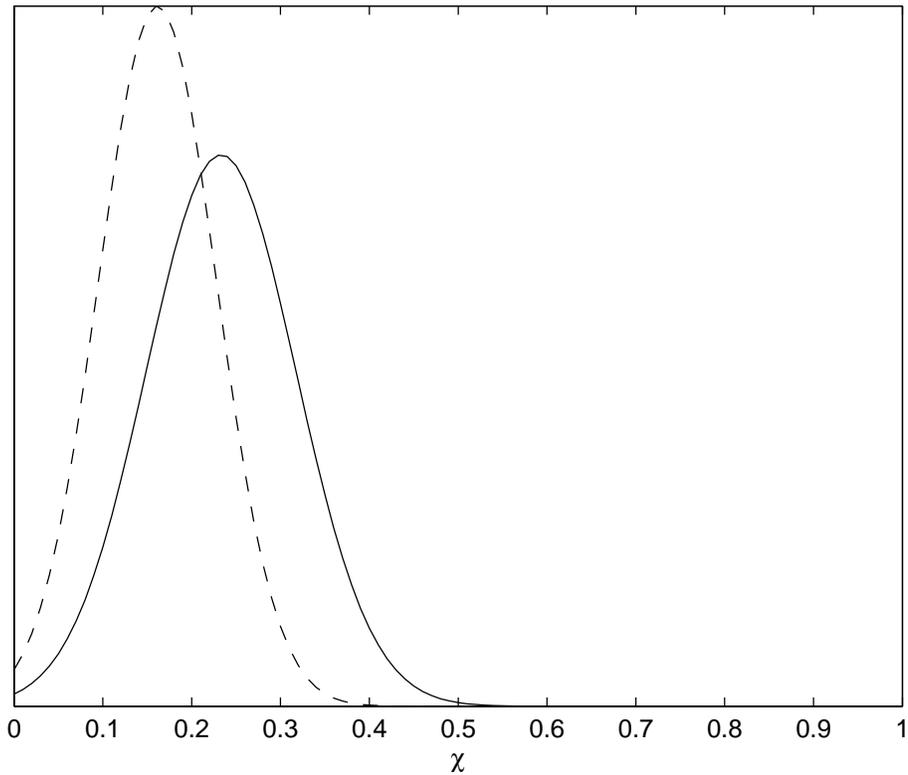}
\caption{Monte-Carlo black-hole spin distribution
following black hole growth via minor mergers from $t=M/m=50$ to $t=M/m=100$.  
The spin distribution for a black hole with initial spin
$\chi=0.9$ is shown with a solid curve, and one
for initial spin $\chi=0.1$ is shown with a dashed curve.\label{Hist50100}}
\end{figure}

In Figure \ref{Hist50100} we plot the spin distribution for a black hole of
mass $t=M/m=100$ that started out at $t=M/m=50$ at either spin $\chi=0.1$ or
$\chi=0.9$ before growing via minor mergers.  This corresponds, for example, to
an intermediate-mass black hole that grows from $70\ M_\odot$ to $140\ M_\odot$
by capturing $M=1.4\ M_\odot$ neutron stars.  The means of the spin at $t=100$
are $\bar{\chi}=0.162$ for the initially slowly-spinning hole and
$\bar{\chi}=0.233$ for the initially rapidly-spinning hole.  The final spin in
the initially rapidly-spinning case decreases as    $\bar{\chi} \sim \chi_0
(t/t_0)^{-2}$,  rather than $\bar{\chi} \sim \chi_0 (t/t_0)^{-2.63}$ as
predicted by Eq.~(\ref{SpinApprox}).  That is because the spin begins to
approach the asymptotic value of  $\bar{\chi} \approx \sqrt{1.5/t} \approx
0.12$ as predicted by Eq.~(\ref{SpinAsympt}), and the rate of spin evolution
decreases because $\chi^2 t$ is no longer much greater than one.   The
initially slowly-spinning case does not quite satisfy $\chi t \gg 1$, so the
Fokker-Planck analysis is suspect; however, Eq.~(\ref{SpinApprox2}), relevant
since $\chi^2 t < 1$ in this case, provides a roughly accurate estimate of spin
growth. The standard deviations at $t=100$ are $\sigma=0.066$ for initial spin
$\chi=0.1$ and $\sigma=0.084$ for initial spin $\chi=0.9$; the predicted
asymptotic value of the standard deviation according to Eq.~(\ref{SigmaAsympt})
is $\sigma=0.087$.  The mass ratios considered in this paragraph may be
plausible for intermediate-mass-ratio inspirals into intermediate-mass black
holes that would be detectable with Advanced LIGO \citep{Mandel2007}.

Finally, we perform a Monte-Carlo simulation of the evolution of a spin
distribution from $t=1100$ to $t=1200$ where the starting mean spin is 
$\bar{\chi}=0.72$ and the starting standard deviation is
$\sigma=0.016$.  In this case, $\chi^2 t \gg 1$ holds throughout the evolution,
so this example can be viewed as a test of our Fokker-Planck analysis.  Based
on Eq.~(\ref{SpinApprox}), we expect the spin  at $t=1200$ to decrease to
$\bar{\chi}=0.57$; in fact, we find $\bar{\chi}
(t=1200)=0.58$.  Since $\sigma^2 t \ll 1$, we expect the standard deviation to
grow via Eq.~(\ref{SigmaApprox2}) to $\sigma=0.022$ at $t=1200$; in fact,
$\sigma(t=1200)=0.021$.

The Fokker-Planck analysis should give excellent results in the regime of very
large $t$, such as those corresponding to minor mergers of stellar-mass compact
objects with $\sim 10^6\ M_\odot$ massive black holes in galactic centers.  (The
extreme-mass-ratio inspirals preceding such minor mergers are an interesting
class of potential LISA sources~\citep{Pau}.)  On the other hand, if a 
large
range of $t$ must be covered, Monte-Carlo simulations become expensive. Thus,
the Monte-Carlo numerical methods and Fokker-Planck analysis can be viewed as
complementary techniques.

\section{Effect of black-hole spin on detection ranges for low-mass-ratio
inspirals} \label{Boost}

The frequency of the last stable orbit before plunge is strongly influenced by
the black-hole spin and the orbital inclination.  Prograde inspirals into
rapidly spinning black holes will have much higher LSO frequencies than
inspirals into non-spinning black holes or polar inspirals into spinning black
holes of the same mass, while retrograde inspirals into rapidly spinning black
holes will have lower LSO frequencies.  For example, for a maximally spinning
Kerr black hole, the frequency of the LSO of a retrograde equatorial inspiral
is twice lower than for a polar orbit, while the LSO frequency of a prograde
equatorial inspiral is six times higher than for a polar orbit.  Even for a
more moderately spinning black hole with $\chi=0.4$, there is almost a factor
of two difference between LSO frequencies for prograde and retrograde
inspirals.

The signal-to-noise ratio (SNR) for the detection of gravitational waves from
inspirals depends on where the LSO frequency falls on the noise power spectral
density curve of the detector.  Although some inclination angles will increase
SNR and others will decrease it, we might generally expect that average
detection range for inspirals into spinning black holes will be higher than
into non-spinning ones.  (``Average'' refers to averaging over the
isotropically distributed orbital inclination angles of the inspiraling
object.)  This is because of the cubic dependence of the detection volume on
detection range, which is proportional to SNR: if, say, $10\%$ of all inspirals
have their SNR boosted by a factor of three, these will be seen three times
further and the detection volume for these kinds of inspirals will go up by a
factor of $27$, so the average volume in which detections can be made will
increase by a factor of $\sim 3$, and the average detection range will grow by
the cube root of $3$.

Conversely, this average detection range increase can manifest itself as a bias
in favor of detecting inspirals into rapidly spinning black holes rather than
slowly spinning ones.  Thus, a numerical estimate of the detection range
increase due to black hole spin is useful for determining whether a high
fraction of rapidly spinning black holes among detected inspirals is an
indication of the prevalence of such black holes in the universe, or whether
this is merely a selection effect.

We use the simple scaling 
\be
|\tilde{h}(f)^2| \propto f^{-7/3}
\ee
for the frequency-domain gravitational wave.  The square of the signal-to-noise
ratio $\rho^2$ is proportional to
\bel{SNR}
\rho^2 \propto 
\int_{f_{\rm min}}^{f_{\rm max}} \frac{|\tilde{h}(f)^2|}{S_n(f)} df \propto
\int_{f_{\rm min}}^{f_{\rm max}} \frac{f^{-7/3}}{S_n(f)} df.
\ee
Here, $S_n(f)$ is the noise power spectral density of the detector, 
$f_{\rm max}$ is the frequency of gravitational waves from the 
last stable orbit, and $f_{m\rm in}$ is the low-frequency cutoff for the
detector for Advanced LIGO, where $f_{\rm min}= 10$ Hz, or the frequency of
gravitational waves one year before plunge for LISA.  We
set $f_{\rm max}$ equal to twice the orbital frequency at the LSO, which we
obtain numerically as a function of the black-hole mass $M$ and spin $\chi$ and
of the orbital inclination angle $\cos{\iota}$ by solving Eq.~(\ref{R}). 

The distance to which an event can be seen is proportional to SNR, $\rho$, so
the detection volume is proportional to $\rho^3$.  Therefore,  we average
$\rho^3$, computed via Eq.~(\ref{SNR}), over the different inclinations
$\cos{\iota}$ (uniformly distributed through the range $[-1,\ 1]$) in order to
compute the expected increase in the detection volume for a given values of
$\chi$, and then take the cube root to compute the increase in the average
detection range. 

\begin{figure}
\includegraphics[keepaspectratio=true,width=6in]{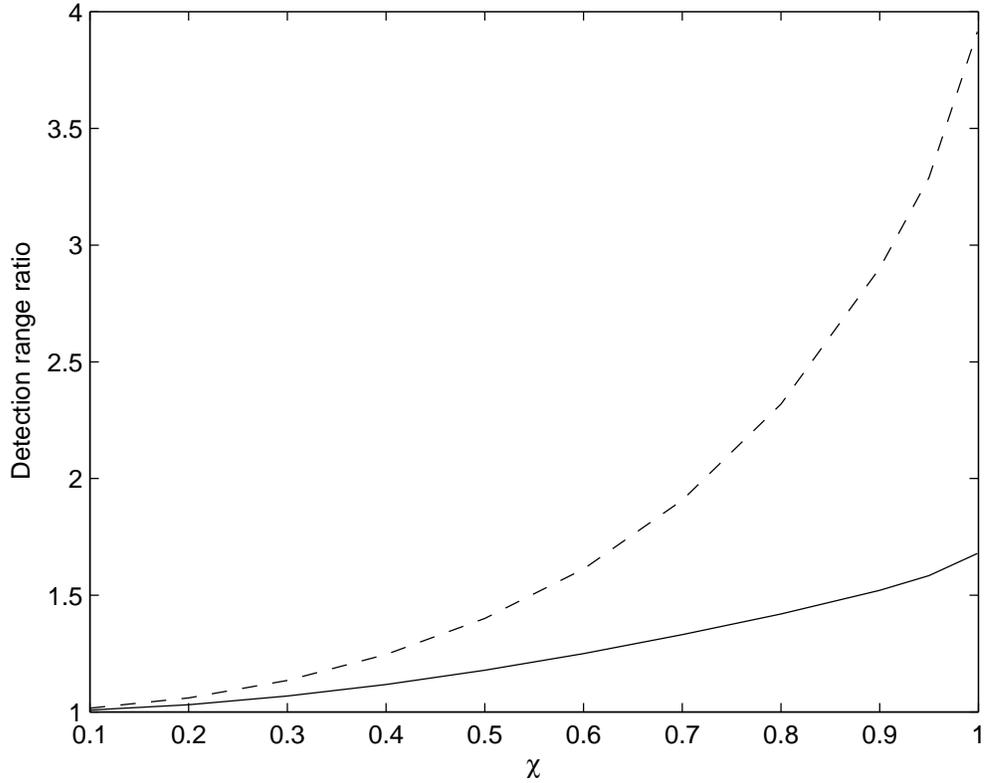}
\caption{The ratio between the inclination-averaged Advanced-LIGO detection
range for intermediate-mass-ratio inspirals into Kerr black holes of a given
spin and the detection range for IMRIs into non-spinning black holes.  The
solid curve represents black holes with mass $M=100\ M_\odot$;  the dashed
curve, mass $M=200\ M_\odot$.
\label{AdvLIGOBoost}}
\end{figure}  

We have computed detection ranges for Advanced LIGO using this method with the
noise power spectral density $S_n(|f|)$ taken from~\citep{Fritschel}. 
Fig.~\ref{AdvLIGOBoost} shows our computed ratio between (i) the average
Advanced-LIGO detection range for intermediate-mass-ratio inspirals into black
holes of a given mass and spin and (ii) the detection range for IMRIs into
Schwarzschild black holes with the same mass. For low spins $\chi \lesssim
0.4$, which are typical for intermediate-mass black holes of $\sim 
100-200$ solar masses that gained a significant fraction of their mass via 
minor mergers, we can approximate the detection range increase due to the 
inclusion of central  black hole spin as 
\bel{AdvLIGORatio} 
\frac{\rm Range_{\rm spin}}{\rm Range_{\rm no-spin}} 
	\sim 1+0.6\chi^2 \left( \frac{M}{100\ M_\odot} \right).
\ee 
This is the ratio of detection ranges; the ratio of detection volumes is a
cube of this ratio.

The effects of cosmological redshift are not significant for Advanced-LIGO
IMRIs when the black-hole spin is small.  Even prograde equatorial inspirals of
neutron stars into $M=100\ M_\odot$ black holes spinning at $\chi=0.9$ are only
detectable to $z \approx 0.2$ at an SNR threshold of $8$.  The cosmological
redshift has the same effect as increasing the black-hole mass, so including
redshift increases the ratio of detection volumes at higher spins.  For the
purposes of including redshift in Fig.~\ref{AdvLIGOBoost}, the inspiraling
object mass was set to $m=1.4\ M_\odot$ and a detection threshold of ${\rm
SNR}=8$ was assumed.    

The results described here do not include higher-order ($m \neq 2$) harmonics
of the orbital frequency.  Higher harmonics are not significant when black-hole
spins are small, since in that case they affect both the spinning and the
non-spinning rates roughly equally, and so the ratio does not change. However,
for high values of spin, the ratios would probably drop somewhat relative to
those given in Fig.~\ref{AdvLIGOBoost}, since including higher-frequency
harmonics would contribute more to increasing the detection range for inspirals
into non-spinning holes than into rapidly holes with prograde orbits
(cf.~Fig.~6 of \citep{Mandel2007}).

We also compute the dependence of the LISA EMRI detection range on the massive
black hole spin.  We consider EMRIs of $m=10\ M_\odot$ objects into $M=10^5\
M_\odot$, $M=10^6\ M_\odot$, and $M=10^7\ M_\odot$ massive black holes.  We
assume that a detection is possible at an SNR threshold of 30.  (Setting the 
threshold to 15 changes the results at the $10-20\%$ level.)  Cosmological
redshift must be included for LISA EMRIs since they can be seen to $z \sim
1-2$.  This means we must specify the inspiraling object mass and the SNR
detection threshold, since these are necessary to  determine the cosmological
redshift of the most distant detectable source.

LISA EMRIs only sweep through a fraction of the frequency band during the
observation time.  Therefore, $f_{m\rm in}$ for LISA is set not by the detector
threshold, but by the frequency of the gravitational waves emitted one year
before plunge.  We compute $f_{m\rm in}$ by evolving the gravitational-wave
frequency back in time from plunge for one year using the prescription of
\citet{BarackCutler2004} (Eqs.~(28) and (29)).

For $M=10^5\ M_\odot$, the spin of the black hole is almost irrelevant: once we
average over orbital inclinations, the spin affects the detection range at a
level of at most a few percent.  This is because at these low masses,  most of
the SNR comes from the portion of the inspiral at much higher radii than the
LSO, so the exact frequency of the LSO does not play a very significant role
(cf.~Fig.~8 and associated discussion in \citep{Pau}).

\begin{figure}
\includegraphics[keepaspectratio=true,width=6in]{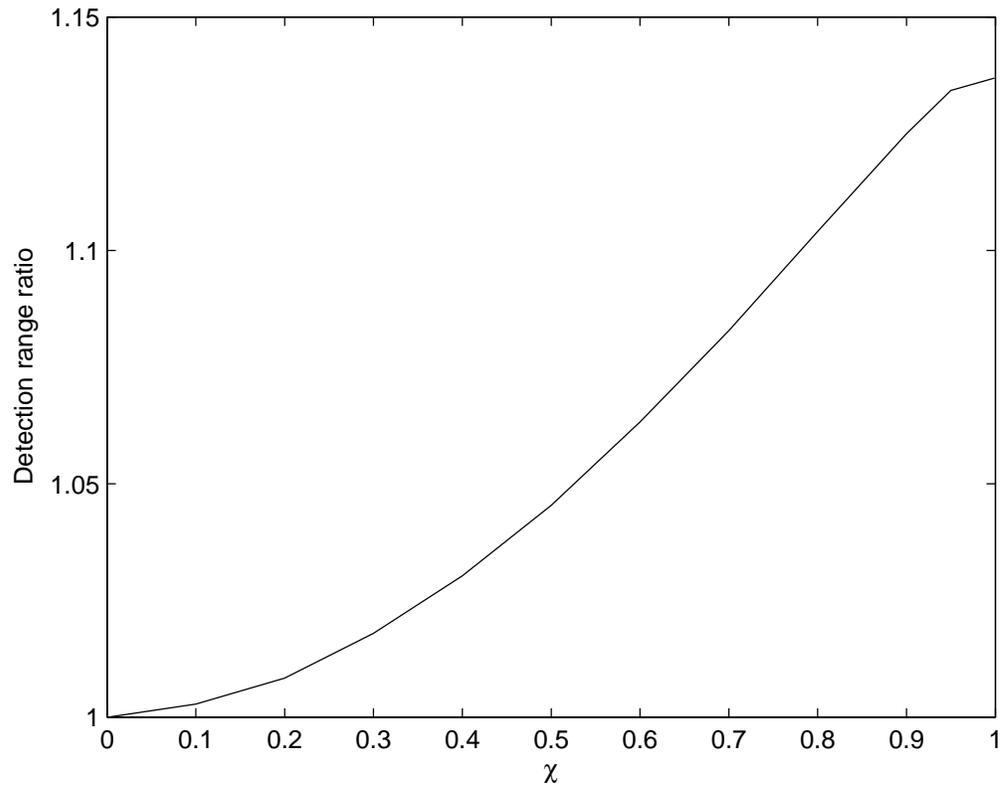}
\caption{The ratio between LISA detection ranges (at SNR$=30$) 
for extreme-mass-ratio inspirals of $m=10\ M_\odot$ compact objects
into Kerr black holes of mass $M=10^6\ M_\odot$ and 
a given spin vs.~non-spinning black holes.   
\label{LISABoost6}}
\end{figure}  

\begin{figure}
\includegraphics[keepaspectratio=true,width=6in]{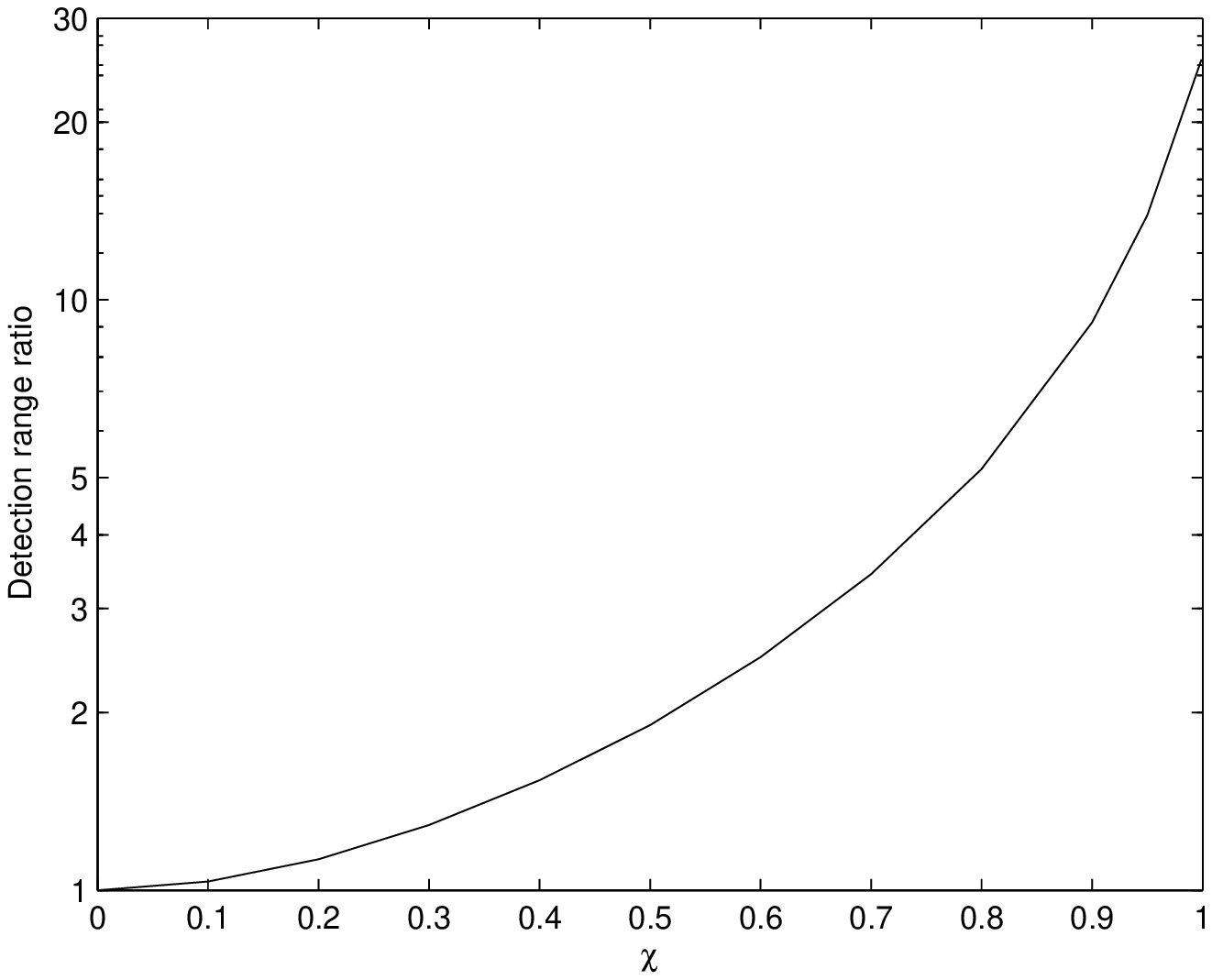}
\caption{The ratio between LISA detection ranges (at SNR$=30$) 
for extreme-mass-ratio inspirals of $m=10\ M_\odot$ compact objects
into Kerr black holes of mass $M=10^7\ M_\odot$ and 
a given spin vs.~non-spinning black holes.   
\label{LISABoost7}}
\end{figure}  

Figure~\ref{LISABoost6} shows the dependence of the average EMRI detection
range on the massive-black-hole spin for $M=10^6\ M_\odot$. The average
detection range for EMRIs into rapidly spinning black holes of mass $M=10^6\
M_\odot$ is $\sim 13\%$ larger than for EMRIs into non-spinning black holes. 
For $M=10^7\ M_\odot$, the detection range for EMRIs into rapidly spinning
black holes is increased by a factor of $\sim 25$ over those into non-spinning
black holes, as shown in  Fig.~\ref{LISABoost7}.  This greater sensitivity to 
black hole spin is expected, since for these massive black holes most of the
SNR comes from the cycles near the LSO.  However, this should not be taken to
mean that inspirals into rapidly spinning $M=10^7\ M_\odot$ black holes are
likely to dominate LISA EMRI observations.  Figures \ref{LISABoost6} and
\ref{LISABoost7} show detection range ratios only; the inclination-averaged
detection range for an EMRI into a maximally spinning $M=10^7\ M_\odot$ black
hole is actually less than the detection range for an EMRI into a non-spinning
$M=10^6\ M_\odot$ black hole.  On the other hand, this large ratio does mean
that there is a strong detection bias in favor of rapidly spinning black holes,
which must be taken into account when statistics of EMRI observations are
inverted to gather information about the massive-black-hole spin distribution.

\begin{acknowledgments}
I thank Kip Thorne and Curt Cutler for suggesting this problem, and Luc
Bouten, Jonathan Gair and Cole Miller for helpful discussions, and Kip Thorne
for thorough comments on the manuscript.  I was partially
supported by NSF Grant PHY-0099568, NASA Grant NAG5-12834, and the Brinson
Foundation.
\end{acknowledgments}


\begin{thebibliography}{}

\bibitem[Barack \& Cutler(2004)]{BarackCutler2004}
{Barack}, L.~\& {Cutler}, C. 2004, \prd {\bf 69}, 082005

\bibitem[Fritschel(2003)]{Fritschel}
Fritschel, P. 2003, arXiv:gr-qc/0308090

\bibitem[G\"ultekin, Miller, \& Hamilton(2004)]{GMH04}
{G\"ultekin}, K., {Miller}, M.~C., \& {Hamilton}, D.~P. 2004,
ApJ, 616, 221

\bibitem[G\"ultekin, Miller, \& Hamilton(2006)]{GMH06}
{G\"ultekin}, K., {Miller}, M.~C., \& {Hamilton}, D.~P. 2006,
ApJ, 640, 156

\bibitem[Hughes(2000)]{Hughes2000}
Hughes, S.~A. 2000, \prd {\bf 61}, 084004

\bibitem[Hughes \& Blandford(2004)]{HB}
Hughes, S.~A. \& Blandford, R.~D. 2004, ApJ, 585, L101

\bibitem[Mandel et al.(2007)]{Mandel2007}
Mandel, I., Brown, D.~A., Gair, J.~R., Miller, M.~C. 
2007, submitted to ApJ, arXiv:0705.0285

\bibitem[Miller(2002)]{Miller2002}
Miller, M.~C. 2002, ApJ, 581, 438

\bibitem[Miller \& Colbert(2004)]{MC2004}
{Miller}, M.~C., \& {Colbert}, E.~J.~M. 2004, IJMPD, 13, 1

\bibitem[Miller \& Hamilton(2002a)]{MH02a}
{Miller}, M.~C., \& {Hamilton}, D.~P. 2002a, MNRAS, 330, 232

\bibitem[Miller \& Hamilton(2002b)]{MH02b}
{Miller}, M.~C., \& {Hamilton}, D.~P. 2002b, ApJ, 576, 894

\bibitem[Misner, Thorne, \& Wheeler(1973)]{MTW}
Misner, C.~W., Thorne, K.~S., \& Wheeler, J.~A. \textit{Gravitation}
(Freeman, San Francisco, 1973)

\bibitem[Mouri \& Taniguchi(2002a)]{MT02a}
{Mouri}, H., \& {Taniguchi}, Y. 2002a, ApJ, 566, L17

\bibitem[Mouri \& Taniguchi(2002b)]{MT02b}
{Mouri}, H., \& {Taniguchi}, Y. 2002b, ApJ, 580, 844

\bibitem[O'Leary et al.(2006)]{OL06}
{O'Leary}, R.~M., {Rasio}, F.~A., {Fregeau}, J.~M., {Ivanova}, N.,
\& {O'Shaughnessy}, R. 2006, ApJ, 637, 937

\bibitem[Amaro-Seoane et al.(2007)]{Pau}
Amaro-Seoane, P., Gair, J.~R., Freitag, M., Miller, M.~C., Mandel, I., 
Cutler, C.~J., Babak, S. 2007, submitted to CQG, arXiv:astro-ph/0703495

\bibitem[Taniguchi et al.(2000)]{Tan00}
{Taniguchi}, Y., {Shioya}, Y., {Tsuru}, T.~G., \& {Ikeuchi},
S. 2000, PASJ, 52, 533

\end{thebibliography}
\end{document}